\documentstyle[psfig,aps,prl,twocolumn]{revtex}

\begin{document}
\draft \wideabs{
\author{Michael K\"ohl, Theodor W.~H\"ansch, and Tilman Esslinger}
\address{Sektion Physik,
Ludwig-Maximilians-Universit\"at, Schellingstr.\ 4/III, D-80799 Munich, Germany and\\
Max-Planck-Institut f\"ur Quantenoptik, D-85748 Garching,
Germany\\ Submitted April 20 2001}
\title{Measuring the temporal coherence of an atom laser beam}
\maketitle

\begin{abstract}
\noindent We report on the measurement of the temporal coherence of an atom laser beam extracted from
a $^{87}$Rb Bose-Einstein condensate. Reflecting the beam from a potential barrier creates a standing
matter wave structure. From the contrast of this interference pattern, observed by magnetic resonance
imaging, we have deduced an energy width of the atom laser beam which is Fourier limited by the
duration of output coupling. This gives an upper limit for temporal phase fluctuations in the
Bose-Einstein condensate.
\end{abstract}

 }
\noindent

One of the fundamental properties characterizing a matter wave source is its degree of temporal
coherence. Perfect coherence in the time domain would allow one to completely predict the phase
evolution of the underlying field. In light optics, a laser comes closest to this ideal situation.
The temporal coherence of a laser exceeds that of a thermal light source by far, which is central to
many applications in spectroscopy, metrology and interferometry. Similarly, a matter wave source
based on Bose-Einstein condensation \cite{bec,atomlasers} is expected to have a substantially higher
degree of temporal coherence than a thermal atom source. So far, experimental investigations of the
coherence of Bose-Einstein condensates have focused on the spatial domain: The interference of two
condensates has been observed \cite{Andrews97}, the uniformity of the spatial phase has been
demonstrated\cite{Hagley99,Stenger99} and the spatial correlation function has been determined
\cite{Bloch00}.

A measurement of the temporal coherence of Bose-Einstein condensates or atom laser beams has not yet
been reported. However, there are prospects to realize matter wave sources with coherence times
comparable to state-of-the-art optical lasers. Theoretically, the energy width of a matter wave beam
extracted from a Bose-Einstein condensate should approach the Fourier limit which is determined by
the duration of the output coupling process\cite{Band99}. Temporal fluctuations of the phase of a
Bose-Einstein condensate are passed on to an atom laser beam that is coherently extracted from the
condensate and will therefore ultimately limit the coherence properties of such beams. Phase
diffusion at finite temperature\cite{Graham98} and fluctuations in the atom number\cite{Wiseman01}
are expected to limit the coherence time of a Bose-Einstein condensate.  The temporal evolution of
the relative phase between two spin components of a Bose-Einstein condensate has been studied
\cite{Hall98}. This measurement has shown the robustness of the relative phase but it was insensitive
to temporal phase fluctuations common to both components of the condensate.

\begin{figure}
\centerline{\psfig{file=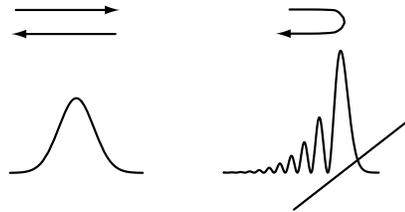,angle=0,width=0.3
\textwidth}} \caption {Principle of the measurement. A wave
packet reflected from a potential barrier develops a standing
wave structure at the turning point.} \label{figure1}
\end{figure}

We investigate the coherence time of an atom laser beam by measuring the contrast of the standing
wave pattern that emerges when the atom laser beam is retro-reflected from a potential barrier (Fig.
\ref{figure1}). This interference process is different from atom optical interference experiments
performed so far, where an atomic wave packet is coherently split and subsequently recombined
\cite{atominterferometry}. In contrast, we study the interference of the reflected front end of the
wave packet with its own back end. The measurement is therefore sensitive to phase fluctuations of
the condensate in the time domain. The atom source and the detection scheme are independent from each
other and common fluctuations are minimal. The reflecting barrier is formed by a linear magnetic
potential several times steeper than the gravitational potential. The spatial structure of the
standing matter wave can not be resolved optically since it is about 1/5 of the $^{87}$Rb resonance
wavelength. We have therefore developed a one-dimensional magnetic resonance imaging method which is
based on RF-spectroscopy between different atomic Zeeman sublevels.

The incoming atom laser beam is prepared in the $|m_F=1\rangle$ Zeeman sublevel of the F=2 hyperfine
ground state and reflected by a magnetic field gradient of $B^{\prime}=200 \ \frac{G}{cm}$. In this
linear potential the stationary solutions of the Schr\"odinger equation are Airy functions
$Ai(\frac{z-z_0}{l})$, where $z_0$ is the apex of the classical trajectory \cite{Landau}. The scaling
parameter $l=(\frac{\hbar^2}{2 m |\frac{dV}{dz}|} )^{1/3}$ is determined by the potential gradient
$\frac{dV}{dz}$ and the mass $m$ of the atom. It has the value $l_{|m_F=1\rangle}=170$ nm for the
magnetic field gradient $B^\prime$ and the atomic state $|m_F=1\rangle$, which has the magnetic
moment $\mu=\mu_B /2$.

\begin{figure}
\centerline{\psfig{file=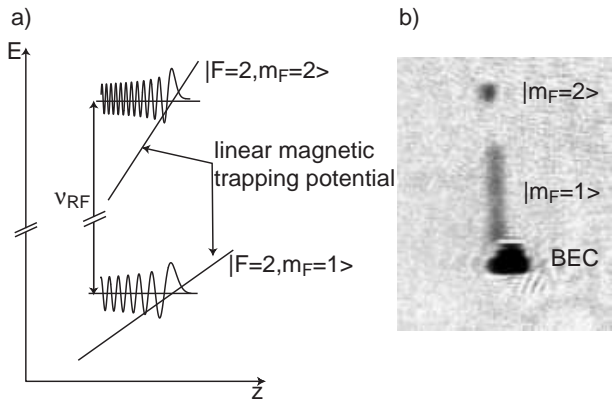,angle=0,width=0.45
\textwidth}} \caption {{\bf (a)} Schematic description of the RF
Spectroscopy. The atom laser approaches the potential barrier in
the $|F=2,m_F=1\rangle$ state. The incident and the
retro-reflected wave form a standing wave pattern. This wave
function is coupled to the $|F=2,m_F=2\rangle$ state by a radio
frequency field. The wave function for atoms in the
$|F=2,m_F=2\rangle$ has different periodicity and the overlap can
be changed by shifting the turning points with respect to each
other. {\bf (b)} Longitudinal Stern-Gerlach separation in the
inhomogeneous magnetic trapping field.} \label{figure2}
\end{figure}

An RF field couples the atoms in the created standing matter wave to the $|m_F=2\rangle$ Zeeman
sublevel which has twice the magnetic moment. In that state the atoms experience approximately twice
the potential gradient and the scaling parameter $l_{|m_F=2\rangle}$ of the  Airy function is
correspondingly smaller (Figure \ref{figure2}a). The transition probability $\it p$ between the two
states is proportional to the overlap integral of the Airy functions
\begin{equation}
p \propto \left | {\int dz \ Ai^*\left (\frac{z-z_{0,|m_F=2\rangle}}{l_{|m_F=2\rangle}}\right ) \
Ai\left ( \frac{z-z_{0,|m_F=1\rangle}}{l_{|m_F=1\rangle}}\right)}\right |^2 .
\end{equation}
The contribution to the integral is significant only where the two functions have similar
periodicity. This is predominantly the case in the vicinity of the turning points. The turning point
of the $|m_F=1\rangle$ atoms is fixed by the total energy of the incoming wave packet. The energy of
the $|m_F=2\rangle$ atoms, and hence their turning point, is set by the RF frequency. A variation in
the RF frequency changes the transition probability $\it p$ since the turning point of the atoms in
the $|m_F=2\rangle$ state is shifted with respect to the turning point of the atoms in the
$|m_F=1\rangle$ state. State selective analysis of the atom laser beam after the reflection allows us
to measure the transition probability $p$ in the experiment.

A Bose-Einstein condensate of $5 \times 10^5$ $^{87}$Rb atoms is created in a QUIC-trap
\cite{Esslinger98} by evaporative cooling in the $|F=1,m_F=-1\rangle$ state. The atom laser beam is
extracted from the condensate using cw output coupling \cite{Bloch99}. A weak, monochromatic RF field
transfers trapped atoms into the $|F=1,m_F=0\rangle$ state where they are accelerated by gravity and
propagate downwards. A collimated beam is formed since the gravitational force largely exceeds the
force that the atom laser beam experiences by the remaining condensate. After a dropping distance of
400 $\mu$m the atoms enter a region of two focused laser beams which transfer all atoms into the
$|F=2,m_F=1\rangle$ state by a two photon Raman transition. The resonance condition for this
transition is given by the difference frequency between the two lasers and the local magnetic field
\cite{Bloch01}. In the low field seeking state $|F=2,m_F=1\rangle$ the atoms experience the
increasing potential of the magnetic trapping field from which the atom laser beam is reflected.
Sufficiently far away from the trap center this potential is given by $V(z)=(\mu B^{\prime} - m g)
z$\cite{Esslinger98}, where $g$ is the gravitational acceleration along the vertical $z$-axis.

Approaching the turning point of their trajectory the atoms are exposed to the $\sigma^+$-polarized
RF field which couples the $|F=2,m_F=1\rangle$ state to the $|F=2,m_F=2\rangle$ state. The fraction
of atoms transferred to this state is determined in the following way. Due to the larger magnetic
moment atoms in the $|F=2,m_F=2\rangle$ state oscillate faster in the magnetic trap and spatially
separate from atoms in the $|F=2,m_F=1\rangle$ state. After half an oscillation period the atoms in
the $|m_F=2\rangle$ state pile up in the upper turning point of their trajectory. At this instant the
magnetic trapping field is switched off and an absorption image is taken from which the peak
absorption of atoms in both states is determined (Figure \ref{figure2}b).

\begin{figure}
\centerline{\psfig{file=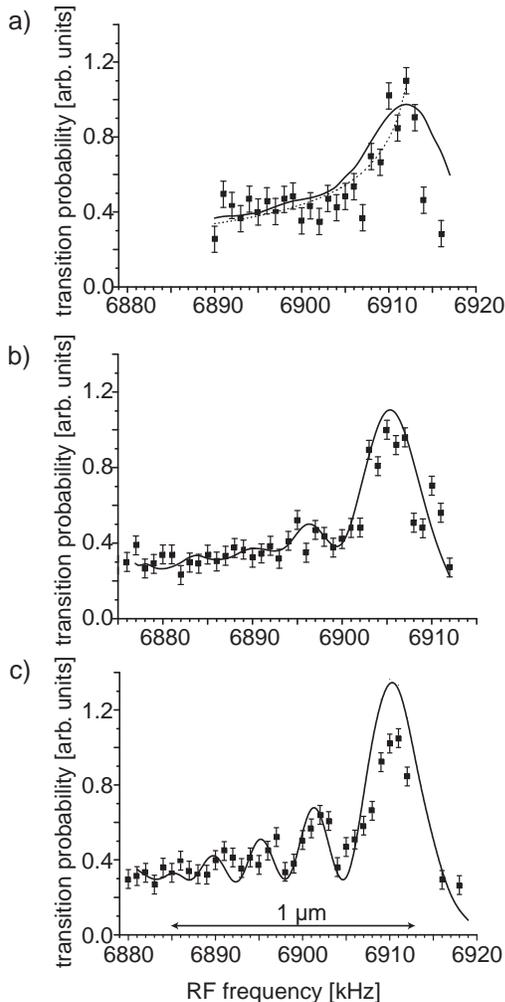,angle=0,width=0.37
\textwidth}} \caption {RF spectra for different output coupling
durations. The overlap integrals (solid lines) are calculated for
an energy width corresponding to the quadratic sum of the Fourier
limited line width and the detector resolution of 1.8 kHz. They
are scaled in amplitude to match the experimental data, but
contain no free parameters. {\bf (a)} 200 $\mu$s atom laser beam.
The dotted line is a fit $\propto 1/\sqrt{z-z_0}$ corresponding
to the classical transition probability. {\bf (b)} 410 $\mu$s
atom laser beam. {\bf (c)} 1.5 ms atom laser beam. The modulation
of the transition probability in (b) and (c) is a signature of
the quantum mechanical character of the reflection process. The
individual data points of the RF spectroscopy have been taken in
different repetitions of the experiment. The error bars are
determined from repetitive measurements at a single frequency.
They are largest for (a) due to the small atom number in the atom
laser beam.  The deviation of the data points from the main peak
in (c) is due to saturation of the RF transition, the slight
mismatch in oscillation frequency is due to fluctuations in the
RF resonance condition, which are discussed in the text. A length
scale is given in (c).} \label{figure3}
\end{figure}

In figure \ref{figure3} RF spectra of standing matter wave pattern are displayed which are taken for
atom laser beams of variable duration. The detected interference pattern directly show the temporal
phase coherence of the wave packet created by the atom laser. The observed contrast increases for
increasing duration of the output coupling process. We compared each data set to a numerical
calculation in which the overlap integral of equation\,(1) is calculated for Airy functions within a
given energy width. We find good agreement with the experimental data when the energy widths for the
calculations are chosen to be the convolution of the Fourier-limit of the output coupling duration
and the detector resolution of 1.8 kHz. The various contributions to the detector resolution are
discussed in detail further below.

For an output coupling period of 1.5 ms we obtain an atom laser linewidth of
$(700^{+400}_{-250})$\,Hz, which is an upper limit for the temporal phase fluctuations of the
Bose-Einstein condensate. The error is obtained from an estimated uncertainty of 10\% in the
convoluted energy width. The energy width of the atom laser beam is smaller than the 2\,kHz
mean-field energy of the condensate and much smaller than the energy span over which output coupling
from the Bose condensate can be achieved, which, due to gravity, is about 15\,kHz \cite{Bloch99}.
Furthermore, we see no evidence that impurity scattering events \cite{Chikkatur00} hinder the
superfluid flow \cite{Raman99} in the output coupling process. Those events would also cause a halo
around the atom laser output, which we do not observe.

The 1.8\,kHz $\pm 0.3$\,kHz energy resolution of our experiment, which corresponds to a spatial
resolution of 65 nm, can be attributed to technical fluctuations and geometrical contributions.

Firstly, there are time-dependent variations of the field strength and position of the magnetic trap.
Short-time fluctuations (5-100 ms) of the magnetic field were minimized by employing a low noise
power supply ($\Delta I_{RMS}/I<10^{-4}$) and by placing the trap inside a magnetic shield enclosure.
The motion of the magnetic trapping coils was passively decoupled from acoustic noise on the optical
table by rubber sockets and the air-conditioning in the laboratory was switched off 10 s before the
atom laser beam was extracted from the condensate. Using a seismic sensor we monitored the vibrations
of the magnetic trap. All described time-dependent fluctuations amount to 700 Hz. Secondly,
shot-to-shot variations of the resonance condition for the two-photon Raman transition modify the
energy with which the atom laser beam approaches the magnetic field gradient barrier. The resonance
condition is determined by the local magnetic field, the  frequency difference between the Raman
lasers, and their intensity. The shot-to-shot reproducibility of the current supply producing the
magnetic field was measured to be better than $6 \times 10^{-5}$ corresponding to 300 Hz. The
difference frequency between the Raman lasers was stabilized to better than 10 Hz. Intensity
fluctuations of the Raman laser beams change the light shift for the two atomic states, but only the
difference in light shift changes the resonance condition. Being detuned $\Delta=70$ GHz from the
$D_1$-line we obtain for our experimental parameters a difference in light shift of 80 kHz/mW, which
we have confirmed experimentally. The intensity of the Raman lasers is actively stabilized to a
relative stability of $3 \times 10^{-3}$ and the detuning was controlled to $\pm$15 MHz by adjusting
current and temperature of the extended cavity diode lasers. Position noise of the Raman lasers with
respect to the magnetic trapping field also changes the intensity of the Raman laser light at the
location of the resonance. We minimize this effect by localizing the spinflip resonance at the center
of the Raman beams and position stabilizing the Raman laser focus with respect to the magnetic
trapping coils. The remaining position jitter of 1/25 of the beam waist results in relative intensity
fluctuations of $3 \times 10^{-3}$ at the center of the focus. The total contribution of technical
noise to the energy resolution is 850 Hz.

The three-dimensional geometry of the magnetic field also limits the energy resolution obtained with
the RF spectroscopy. Away from the center of an elongated Ioffe trap there is a weak axial magnetic
field gradient, transverse to the atom laser beam. Therefore atoms on one side of the beam are
reflected at a slightly different height compared to atoms on the other side. For a diameter of the
atom laser of 70 $\mu$m this amounts to 2 kHz energy difference across the beam. By evaluating the
optical density in the absorption images only in the center of the reflected wave packets we can
reduce this effect by a factor of 4. The resonance condition for the RF spectroscopy is given by the
surface of constant magnetic field strength. In the radial plane this resonance shell is misaligned
with respect to the surfaces of constant energy of the reflection barrier. This misalignment is due
to gravity and amounts to 1.5 kHz across the beamwidth.

From an atom optical point of view the magnetic trapping
potential is a matter wave cavity for the atoms in the
$|F=2,m_F=1\rangle$ state. The observed interference fringes
unambiguously show the spatial structure of the modes in this
cavity. The formation of a standing wave pattern demonstrates
that the cavity "mirrors" \cite{Bloch01} preserve the coherence
of the incident atoms. The longitudinal mode spacing of the
cavity for our parameters is $\Delta \omega = 2 \pi \times 63$Hz
which means that we populate about 10 modes in the experiment,
depending on the output coupling duration. This is an improvement
of three orders of magnitude over previous experiments with laser
cooled atoms, where the number of populated modes is determined
by the size and temperature of the cold atom source. With a
further enhancement of the energy resolution in our experiment it
should be possible to manipulate individual modes in a matter
wave cavity. An alternative experiment towards the observation of
a standing matter wave in a linear potential is proposed for
ultracold neutrons \cite{Nev2000}.

The experimental resolution of the RF spectroscopy may be improved by focusing the atom laser beam
onto the reflecting magnetic field gradient. The geometrical energy width scales approximately
linearly with the beam diameter, so focusing by a factor of 100 will greatly improve the resolution.
It seems feasible to achieve a resolution of a few ten Hz by reducing the light shift fluctuations
and enhancing the magnetic field stability when operating the coils from a battery. In this regime a
transition to a Lorentzian lineshape of the atom laser is expected, when output coupling rate (which
is much smaller than the trapping frequencies $\Gamma \simeq 10 $s$^{-1}\ll \omega=2 \pi \times$100
Hz) dominates compared to the output coupling duration \cite{Band99}. For a reduced output coupling
rate a decrease of the coherence time due to number fluctuations in the condensate, which is expected
to be on the order of ten Hz\cite{Wiseman01}, might become visible. Analogous to the Schawlow-Townes
limit for optical lasers \cite{Schawlow58}, phase diffusion processes \cite{Graham98} will ultimately
limit the linewidth of an atom laser to a few Hz.

In conclusion, we have measured the temporal coherence of an atom laser beam. A standing matter wave
is created by retro-reflecting the atom laser beam from a potential barrier. Employing magnetic
resonance imaging we detect the interference structure with a spatial resolution of 65 nm. For the
atom laser beam we deduce a Fourier limited energy width of 700 Hz, which is substantially below the
mean field energy of the Bose-Einstein condensate. Our results show that phase fluctuations in the
condensate are negligible on the time scale of our measurement and that the output coupling process
preserves the coherence of the atom laser.

We would like to thank R. de Vivie-Riedle for discussions, H. Gebrande for the loan of a seismic
sensor, C. Kurtsiefer for the loan of a high voltage amplifier, and DFG for financial support.

\end{document}